\documentclass[twocolumn,showpacs,nofootinbib,amsmath,amssymb,aps,prd,superscriptaddress]{revtex4-1}

\pdfoutput=1

\usepackage[CJKbookmarks, bookmarksnumbered, bookmarksopen, colorlinks, citecolor=blue, linkcolor=blue]{hyperref}
\usepackage[english]{babel}
\usepackage{amstext,amsmath,amssymb,amsfonts,bbm}
\usepackage[latin1]{inputenc}
\usepackage{graphicx}
\usepackage{bmpsize}
\usepackage{epstopdf}
\usepackage{amsthm}
\usepackage{tocvsec2}
\usepackage{enumerate}
\usepackage{color}
\usepackage{tikz}

\usepackage{braket}

\newcommand{\va}{\scriptscriptstyle}

\newcommand{\be}{\nopagebreak[3]\begin{equation}}
\newcommand{\ee}{\end{equation}}

\newcommand{\bee}{\nopagebreak[3]\begin{equation*}}
\newcommand{\eee}{\end{equation*}}

\newcommand{\ba}{\nopagebreak[3]\begin{eqnarray}}
\newcommand{\ea}{\end{eqnarray}}
\DeclareFontFamily{U}{rsfs}{}         
\DeclareFontShape{U}{rsfs}{m}{n}{<5> rsfs5 <6><7> rsfs7          %
  <8><9><10><10.95><12><14.4><17.28><20.74><24.88> rsfs10}{}     %
\DeclareMathAlphabet{\mathfs}{U}{rsfs}{m}{n}                     %
\newcommand{\n}{{\nonumber}}

\newcommand{\f}{\frac}
\newcommand{\extd}{{\rm d}}

\usepackage{color}

\begin{document}

\title{Resolving the $H_0$ tension with diffusion}

\author{Alejandro Perez}
\email{alejandro.perez@cpt.univ-mrs.fr}
\affiliation{{Aix Marseille Univ, Universit\'e de Toulon, CNRS, CPT, Marseille, France}}

\author{Daniel Sudarsky}
\email{sudarsky@nucleares.unam.mx}
\affiliation{Instituto de Ciencias Nucleares,
Universidad Nacional Aut\'onoma de M\'exico, M\'exico D.F. 04510, M\'exico}

\author{Edward Wilson-Ewing}
\email{edward.wilson-ewing@unb.ca}
\affiliation{Department of Mathematics and Statistics,
University of New Brunswick, Fredericton, NB, E3B 5A3, Canada}


\begin{abstract}
The tension between the value of the Hubble constant $H_0$ determined from local supernovae data and the one inferred from the cosmic microwave background based on the $\Lambda$CDM cosmological model may indicate the need for new physics. Here, we show that this `Hubble tension' can be resolved in models involving an effective energy flux from the matter sector into dark energy resulting naturally from a combination of unimodular gravity and an energy diffusion process. The scheme is one where dark energy has the standard equation of state $w=-1$. This proposal provides an alternative phenomenological paradigm accounting for the observations, while offering a general framework to study diffusion effects coming from novel fundamental physical processes.
\end{abstract}

\maketitle

\section{Introduction}

Over the last 30 years there has been a spectacular development of precision cosmology, with the state of the art including high resolution observations of the cosmic microwave background (CMB) by the Planck collaboration \cite{Adam:2015rua, Aghanim:2018eyx}, detailed studies of baryon acoustic oscillations by BOSS \cite{Alam:2016hwk}, and extensive supernovae observations \cite{Riess:2019cxk, Reid:2019tiq, Freedman:2019jwv}. This trend is expected to continue in the near future with information coming from gravitational wave observations by the LIGO collaboration \cite{Farr:2019twy}, as well as the data from ongoing and future missions like GAIA, EUCLID, and the James Webb Space Telescope.

While these observations are, for the most part, consistent with the $\Lambda$CDM concordance cosmology based on a still mysterious dark sector including cold dark matter (CDM) and a cosmological constant $\Lambda$, the increased precision has brought with it further problematic aspects to our current understanding.
 
Dark matter was introduced phenomenologically in order to account for local gravitational effects, as observed in the rotation curves of galaxies, and to offer a viable paradigm for structure formation \cite{Bergstrom:2000pn}. However, it remains unclear whether dark matter is indeed simply an invisible contribution to the matter content of the universe, or an effect due to modifications of general relativity \cite{Clifton:2011jh}. Furthermore, while dark energy is perhaps most simply explained by a cosmological constant, from a theoretical perspective it remains unclear why its value is so small (and, in particular, why quantum vacuum fluctuations do not appear to contribute as naively expected \cite{Martin:2012bt}). Dark energy is estimated to contribute about $70\%$ to the mean energy density of the universe today, as first extracted from supernovae measurements \cite{Riess:1998cb, Perlmutter:1998np}. The situation has motivated numerous phenomenological alternatives to a cosmological constant, such as quintessence and modified gravity theories \cite{Clifton:2011jh}. However, it is safe to say that no fully satisfactory fundamental explanation of these issues seems available at the moment. 

More recently, a statistically significant tension has grown between the values of the Hubble expansion rate today, $H_0$, extracted from supernovae data, and the one inferred from the CMB observations based on the $\Lambda$CDM model \cite{Bernal:2016gxb, Freedman:2017yms}, with a higher value for $H_0$ preferred by supernovae data. This discrepancy is currently reported to be at the level of $4.2\sigma$, for studies based on supernovae observations calibrated using Cepheid stars \cite{Riess:2019cxk, Reid:2019tiq} (although the tension is significantly reduced when using a different distance ladder calibration based on stars lying on the tip of the red giant branch \cite{Freedman:2019jwv}). Could this $H_0$ tension be offering a clue concerning some unknown aspect about the fundamental nature of gravity and matter? 

Various proposals have been put forward in order to alleviate this tension. These typically involve modifying the standard $\Lambda$CDM cosmology in its dark sector: for example, dark radiation \cite{Bernal:2016gxb}, early dark energy \cite{Poulin:2018cxd} or dynamical dark energy \cite{Colgain:2018wgk}, and interacting dark matter/energy models \cite{Kumar:2019wfs, DiValentino:2019ffd, Agrawal:2019dlm}. The tension can also be alleviated (although only partially) for example either by allowing for the variation of Newton's constant (and post-Newtonian parameters) \cite{Rossi:2019lgt}, or by allowing for a more general equation of state ($w\neq-1$) for the dark energy sector, with the data preferring the inclusion of regimes where $w<-1$ \cite{Aghanim:2018eyx}. From a theoretical viewpoint, the last possibility seems rather problematic, as it would imply the violation of all energy conditions by the dark energy. Particularly worrying is the violation of the dominant energy condition, because, in any field theoretical model, that would imply the existence of a tachyonic degree of freedom and acausal energy fluxes.

Recently, a new mechanism has been proposed that generates an effective cosmological constant out of small departures from the strict conservation of energy momentum in the matter sector \cite{Josset:2016vrq}. Although within the context of general relativity violations of the conservation of the stress-energy tensor are inconsistent, a slight modification of the theory known as unimodular gravity permits a specific kind of violation generating a time-dependent $\Lambda$ in the resulting effective Einstein equations \cite{Josset:2016vrq}. This framework must, of course, be complemented with detailed input concerning the form and magnitude of the violations of energy-momentum conservation. If such violations are due to a diffusive process resulting from the space-time micro-structure or granularity in the fabric of space-time at the Planck scale, arising in conjunction with curvature (and therefore absent in the Minkowski space-time \cite{Collins:2004bp, Collins:2006bw}), the result is a cosmological constant of the order of magnitude of its observed value \cite{Perez:2017krv, Perez:2018wlo}. It seems, therefore, natural to consider whether a related mechanism could be at play in late cosmological times leading to an effective modification of the cosmological evolution that could resolve the $H_0$ tension.
 
A simple analysis indicates that the specific model considered in \cite{Perez:2017krv, Perez:2018wlo} is not able to account for any significant modification of $\Lambda$ in the late universe. However, \cite{Perez:2019gyd} considers a related diffusion mechanism, connected in this case with the physics of black holes, which naturally places its occurrence after decoupling, and thus the relevant time period for a possible resolution of the $H_0$ discrepancy.

The purpose of the present work is to explore, in a rather general manner, the possibility that a process taking place at relatively late times, involving an effective violation of local energy conservation in the matter sector within the context of unimodular gravity, might resolve the $ H_0$ discrepancy; due to the resulting growth of the dark energy at post-CMB times changing the late cosmic evolution. Such models could be regarded, at the phenomenological level, as versions of some models with interacting matter and dark energy such as those considered in \cite{Kumar:2019wfs, DiValentino:2019ffd, Agrawal:2019dlm}, but the perspective offered by the unimodular framework implies the origin of the interaction comes from a diffusive process (see also \cite{Banerjee:2019kgu}).

The investigation of a possible diffusion mechanism is opened by the perspective adopted here. A possible model is proposed in \cite{Perez:2019gyd} where the rotational energy of black holes is dissipated via friction produced by fundamental space-time granularity. The details of the energy diffusion during cosmological evolution is determined by the fundamental diffusion equation for each black hole $\dot E= F(M, J)$ together with the expression for the number density of black holes as a function of their mass, angular momentum, and the cosmic time $n(M, J, t) $. In this model the detailed analysis of its effects necessitates the theoretical modelings of the cosmic evolution of black hole abundances. The viability of these models will be constrained in part by astrophysical bounds on the envisioned energy dissipation.

The outline of the paper is the following: in Sec.~\ref{s.ug}, we review the framework of unimodular gravity that allows for the effective energy transfer from the matter to the dark energy sector. In Sec.~\ref{s.tension}, we review the main observational inputs that lead to the $H_0$ tension and their dependence on the cosmological model. In Sec.~\ref{s.models}, we consider some phenomenological models that can resolve the $H_0$ tension. We then offer, in Sec.~\ref{s.int-z}, a brief recount of some intermediate red-shift observations that seem to favor some general features of the phenomenological models considered in this work, and we end with a discussion in Sec.~\ref{s.disc}. To avoid confusion, in the remainder of the paper we will refer to the standard cosmological model as GR-$\Lambda$CDM.

\section{Unimodular gravity and violations of energy momentum conservation}
\label{s.ug}

General Relativity is incompatible with any violation of the conservation of the stress-energy tensor as a consequence of the (contracted) Bianchi identities of the Einstein tensor, i.e., $\nabla^a G_{ab} = 0$. There is, however, a simple modification of general relativity, known as unimodular gravity (UG) \cite{Ellis:2010uc}, introduced by Einstein himself in 1919, that is more permissive in this regard. The field equations of UG are simply the trace-free part of the Einstein equations, namely
\be\label{TFE}
R_{ab} - \frac{1}{4} R g_{ab} = {8 \pi G} \left( T_{ab} - \frac{1}{4} T g_{ab} \right),
\ee
which can be written in the more convenient form
\be\label{TFE1}
R_{ab} - \frac{1}{2} R g_{ab}+\frac{1}{4}\left( R+ {8 \pi G} T \right) g_{ab} = {8 \pi G} T_{ab}.
\ee

Unimodular gravity can be derived from the Einstein-Hilbert action by restricting the variational problem to variations that preserve the (four) volume-form, $g_{ab}\delta g^{ab}=0$ (alternatively, it is possible to add a constraint to the action that requires the volume element of the metric to have a specific value). This is arguably the simplest modification of gravity that completely trivializes the problem of the large contributions to the cosmological constant from vacuum fluctuations in quantum field theory by fully decoupling vacuum energy from the dynamics of the metric \cite{Weinberg:1988cp}.   

As a side remark, note that such a modification could be natural if at a more fundamental level, as the full quantum gravity regime is approached, there are space-time related structures not fully characterized by the metric description and the metric corresponds to some sort of `mean field' or effective description of space-time, perhaps due to a fundamental granularity or a four-volume distribution of singular quantum events. Indeed, unimodular gravity is the gravitational theory emerging from Sorkin's causal set approach \cite{Sorkin:2003bx} and also from the thermodynamical arguments given by Jacobson \cite{Jacobson:1995ab, Jacobson:2015hqa}.

Returning to the trace-free Einstein equations, taking the divergence of \eqref{TFE1} and using the Bianchi identities one finds
\begin{equation}\label{BianchiUG}
\frac{1}{4}\nabla_a\left( R+ {8 \pi G} T \right) = {8 \pi G} \nabla^b T_{ab} \equiv J_a,
\end{equation}
where we have introduced the current of energy momentum violation $J_a = 8 \pi G \nabla^b T_{ab}$. The invariance of unimodular gravity under volume-preserving diffeomorphisms implies that $\extd J=0$ \cite{Josset:2016vrq}, and therefore \eqref{BianchiUG} can be integrated, leading to 
\begin{equation}\label{TFE2}
R_{ab} - \frac{1}{2} R g_{ab} +\left(\Lambda_{0} + \int_{\ell} J\right) g_{ab} = 8 \pi G T_{ab} ,
\end{equation}
where $\Lambda_0$ is a constant of integration and the `energy violation current' $J$ is integrated along any arbitrary path $\ell$ from some reference event to the point where the equation is evaluated. Clearly, this generates an effective $\Lambda(x^a) = \Lambda_0 + \int_\ell J$ which depends on the space-time point. The independence of $\Lambda(x^a)$ on the choice of path $\ell$ is guaranteed by the integrability condition $\extd J=0$.  An analogous system of equations can also be obtained by allowing for certain types of interactions between scalar-field dark energy and the matter fields in Einstein gravity \cite{Benisty:2017eqh, Benisty:2018oyy}.

In the case that the stress-energy tensor is conserved, $J=0$ and \eqref{TFE2} reduces to the Einstein equations, with the integration constant $\Lambda_{0}$ becoming the cosmological constant. However, in contrast to general relativity, the energy-momentum conservation $\nabla^b T_{ab} = 0$ does not follow directly from the equations of motion, and, in fact, it is usually postulated as an additional assumption. This assumption can be replaced by the demand that the matter action be diffeomorphism invariant (related to the assumption that space-time geometry and matter fields are smooth to all scales). For our purposes, the central observation is that, in the context of UG, this extra assumption can be relaxed, leading to a changing cosmological `constant' $\Lambda(t)$.

There are, indeed, various reasons to consider the possibility that $\nabla^b T_{ab} \neq 0$. First of all, the classical setting, where general relativity is typically applied, can be nothing more than a very good approximation. At some point, a quantum description of matter is required and that implies that, at a minimum (in the so-called semiclassical approximation), $T_{ab} $ must be replaced by $\langle \hat T_{ab} \rangle $ (the expectation value of the (renormalized) energy momentum operator in a suitable state). The viability of a semiclassical approximation has been a subject of substantial debate and controversy \cite{Eppley1977, Page:1981aj, Carlip:2008zf}. In fact, such considerations seem to connect, to a large extent, to conceptual difficulties inherent to quantum theory. One of these difficulties is that none of the currently proposed approaches addressing the measurement problem allows for the conservation of energy-momentum at the semi-classical level \cite{Maudlin:2019bje}. 

Another motivation for $\nabla^b T_{ab} \neq 0$ comes from ideas intimately tied to quantum gravity. While there does not yet exist a completely satisfactory quantum theory for gravity, many approaches to the subject predict the existence of some sort of space-time discreteness, in which case the rationale for $\nabla^b T_{ab} = 0$ (smoothness at all scales or diffeomorphism invariance) is lost. This opens the door for what could be naively interpreted as an `energy diffusion' resulting from the interaction of the matter sector and the microscopic structure of space-time \cite{Perez:2017krv, Perez:2018wlo}.  

It is interesting to point out that in a standard quantum mechanical context one should also expect diffusion to be associated with decoherence with the Planckian microscopic structure. This would lead to an apparent violation of unitarity from the perspective of the effective treatment, and thus has been shown to be potentially useful for the resolution of the information loss paradox in the context of black hole evaporation \cite{Amadei:2019ssp, Amadei:2019wjp, Perez:2014xca, Perez:2017cmj}. A contrasting picture is offered by spontaneous collapse modifications of quantum mechanics where unitarity and energy conservation are broken fundamentally.  Such an alternative could be helpful in dealing with certain difficulties faced by the current paradigm of structure formation in cosmology \cite{Perez:2005gh, Leon:2017sru} as well as in resolving the black hole information loss problem in a different way \cite{Okon:2013lsa, Okon:2014dpa, Modak:2014vya, Bedingham:2016aus,Okon:2017pvc}. Such modifications of the quantum theory exist in versions that can both lead to energy diffusion as well as energy increase \cite{Bassi_2005}.

For the remainder of the paper, we will avoid further discussion of the possible fundamental source of the violation of $\nabla^b T_{ab} = 0$, and we will rather focus on the phenomenology of $\nabla^b T_{ab} \neq 0$ in unimodular cosmology, and especially on its possible effects in the context of the $H_0$ tension.

As noted in \cite{Josset:2016vrq}, the homogeneous and isotropic Friedman-Lema\^itre-Robertson-Walker cosmological setting is one where the integrability condition $\extd J = 0$ is automatically satisfied as a result of the assumed symmetries. We will consider here the spatially flat case corresponding to the metric: 
\begin{equation}
\extd s^2 = - \extd t^2 + a^2(t) \, \extd \vec x^2,
\end{equation}
where the expansion rate of the universe is given by the Hubble rate $H = \dot{a}/a$, with the dot denoting a derivative with respect to cosmic time $t$. For simplicity, we choose the normalization of the scale factor $a(t)$ so that its value today is $a(t_0)=1$. Due to spatial homogeneity, the current $J_a$ can only have a non-trivial $(\extd t)_a$ component, and can only depend on $t$, thus we write
\be
J = \dot \Lambda(t) \, \extd t.
\ee

The Friedman and Raychaudhuri equations for unimodular gravity follow directly from \eqref{TFE2}: they are exactly the same as in general relativity, except that $\Lambda$ is no longer a constant:
\be \label{ug-fr}
H^2 = \f{8 \pi G}{3} \sum_i \rho_i + \f{\Lambda(t)}{3}, \quad \dot H = - 4 \pi G \sum_i(\rho_i + p_i),
\ee
where the index $i$ denotes the different matter fields in the space-time. It is often convenient to rewrite the Friedman equation as $1 = \sum_i \Omega_i + \Omega_\Lambda$, with $\Omega_i = 8\pi G \rho_i/3H^2$ and $\Omega_\Lambda = \Lambda/3H^2$.

The growth in $\Lambda$ resulting from the energy diffusion in the matter sector is governed by $\dot\Lambda = 8\pi G \nabla^a T_{ta}$. Given that we are mostly interested in late-time (post-CMB) physics, and at late times the contribution from radiation to the Friedman equation is negligible, we will focus our attention on the energy density $\rho_m$ of dark matter and baryonic matter. That is, we will make the assumption that the growth in $\Lambda(t)$ is entirely due to diffusion from $\rho_m$ (a physical motivation for this exists within the framework of \cite{Perez:2017krv, Perez:2018wlo}). As is usual in cosmology, we will work in the approximation in which matter is taken to have vanishing pressure, and thus the modified continuity equation for $\rho_m$ is simply
\be \label{ug-continuity}
\dot \rho_m + 3 H \rho_m = -\frac{\dot \Lambda(t)}{8\pi G}.
\ee
Expressing everything in terms of the redshift variable $z \equiv (1-a)/a$, this can be rewritten as
\be \label{difu}
(1+z)^3\frac{\extd}{\extd z}\left(\frac{\rho_m(z)}{(1+z)^3}\right)=-\frac{1}{8\pi G}\frac{\extd {\Lambda(z)}}{\extd z}.
\ee

Specific solutions can be found, for example, by considering in detail a particular diffusion process as done in \cite{Perez:2017krv, Perez:2018wlo}, or by proposing a phenomenological model for $\Lambda(z)$ which can be integrated to find $\rho_m(z)$, or vice versa. In the present work we will adopt the second strategy.

\section{The $H_0$ tension in standard and modified models }
\label{s.tension}

The present value of the Hubble parameter $H_0 = H(t_0)$ can be measured by different methods.  One of the most direct and accurate ways to do it is by studying the magnitude of type Ia supernovae and their redshift $z = (1-a)/a$. Inferring the luminosity distance $d_L$ of each supernova from its apparent and absolute magnitudes gives a relation $d_L(z)$. In general, we might write $d_L(z) $ if we have the value of $H(z')$ for $ z'$ in the intervening cosmological regime between the light emission at $z$ and its detection at $ z=0$, namely:
\be \label{luminosity-distance}
d_L = (1+z) \int^{z}_{ 0} H(z')^{-1} \extd z'. 
\ee
If we just want to focus on small values of the redshift $z$, we could instead consider a Taylor expansion of the scale factor $ a(t)$ as a function of cosmic time and write
\be \label{taylor-a}
a(t) = 1 + H_0 \Delta t - \frac{q_0}{2} H_0^2 (\Delta t)^2 + \frac{j_0}{6} H_0^3 (\Delta t)^3 + \ldots
\ee
where $\Delta t = t-t_0$, $q_0$ is the deceleration and $j_0$ is the jerk parameter (and recall that $a_0=1$). A short calculation shows that, for small $z$, the relation between the luminosity distance and the redshift is \cite{Visser:2003vq, Aviles:2012ay}
\be
d_L = \frac{z}{H_0} \Bigg[ 1 + \frac{1 - q_0}{2} z - \frac{1 - q_0 - 3 q_0^2 + j_0}{6} z^2 + \ldots \Bigg].
\ee
The best fit to this relation for type Ia supernovae data for redshifts $z \lesssim 0.15$ gives $H_0 = 73.5 \pm 1.4 ~ {\rm km \, s^{-1} \, Mpc^{-1}}$ \cite{Reid:2019tiq}, based on the Cepheid distance ladder; the tip of the red giant branch distance ladder gives a lower value, $H_0 = 69.8 \pm 2.5 ~ {\rm km \, s^{-1} \, Mpc^{-1}}$ \cite{Freedman:2019jwv}. For observations at small $z$, the observational constraints on $j_0$ and higher order terms are weak, and the determination of $H_0$ is relatively insensitive to uncertainties in $q_0, j_0$, etc. 

Concrete cosmological models correspond to specific values of such parameters. For instance, a spatially flat GR-$\Lambda$CDM cosmology with $\overline \Omega^0_m = 0.3$ and $\overline \Omega^0_\Lambda = 0.7$, corresponds to $\overline q_0=-0.55$ and $\overline j_0=1$ (see Sec.~\ref{s.int-z}); we put bars on these parameters to denote that they correspond to the ones extracted from the data using the GR-$\Lambda$CDM flat model, see below.

Given a particular cosmological model, it is also possible to infer the value of $H_0$ via observations of the CMB. For instance, assuming a spatially flat GR-$\Lambda$CDM cosmology, comparing the CMB data to the theoretical predictions calculated by a Boltzmann code gives $H_0 = 67.4 \pm 0.5 ~ {\rm km \, s^{-1} \, Mpc^{-1}}$ \cite{Aghanim:2018eyx}. There is a $\sim 10\%$ difference compared to the supernovae data based on the Cepheid distance ladder that is statiscally significant at a level of $4.2\sigma$ \cite{Reid:2019tiq}, although the tension decreases when the analyis is carried out using the tip of the red giant branch distance ladder \cite{Freedman:2019jwv}.

We can gain some insight into how $H_0$ can be inferred from the CMB without having to rely on a full Boltzmann code in which many other effects are considered simultaneously, by considering a simpler (and yet nonetheless accurate) calculation. The angular location of the acoustic peaks in the CMB has been measured extremely accurately, and the extracted angular scale $\theta$ provides an excellent approximation to the ratio of the radius of the sound horizon $r_{\rm s}$ (the distance sound waves can travel from the time of reheating to recombination, which is entirely determined by pre-CMB physics) to the radius of the surface of last scattering $R_{\rm \va LS}$ (entirely determined by post-CMB physics),
\be\label{titi}
\theta \approx \frac{r_{\rm s}}{R_{\va \rm LS}},
\ee
with $R_{\va \rm LS}$ given by
\be\label{Rr}
R_{\va \rm LS} = \int_{t_{ls}}^{t_0} \frac{\extd t}{a(t)} = \int_{z_0=0}^{z_{\va \rm LS}=1090} \frac{\extd z}{H(z)},
\ee
using $\extd t = (a H)^{-1} \extd a = -a \, \extd z$. A given cosmological model determines $H(z)$ via the Friedman equation, and that can be used in the expression above to evaluate $R_{\va \rm LS}$. For the GR-$\Lambda$CDM model,
\be
H(z)^2 = {\overline H}_0^2 \Big[ \overline\Omega_r^0 (1+z)^4 + \overline\Omega_m^0 (1+z)^3 + \overline \Omega_\Lambda^0 \Big],
\ee
where $\overline H_0$, $\overline \Omega_i^0$ are the value of the Hubble parameter and the contributions to critical density today in the GR-$\Lambda$CDM model. Although both $\overline\Omega_r^0$ and $\overline\Omega_m^0$ are relevant in the previous equation because radiation is an important contribution close to the CMB time, the calculations that follow will evaluate the Friedman equation at $z=0$ where the contribution from radiation can be neglected. For that reason we only need the explicit value of $\overline\Omega_m^0$ \cite{Aghanim:2018eyx},
\be \label{lilo}
\overline\Omega_m^0=0.315 \pm 0.007,
\ee
where quantities with a bar refer to the GR-$\Lambda$CDM concordance model. If $\theta$ and $r_{\rm s}$ are known then by combining \eqref{titi} and \eqref{Rr}, in the GR-$\Lambda$CDM model $\overline H_0$ is given by
\be
{\overline H_0=\frac{\theta}{r_s} \int\limits_{0}^{z_{\rm LS}} \frac{\extd z}{\sqrt{\overline \Omega_r^0 (1+z)^4 + \overline \Omega_m^0 (1+z)^3 + \overline\Omega_\Lambda^0}}}.
\ee 

In the unimodular gravity type of models that we will consider here, the Friedman equation will be modified. However, these modifications occur at redshift values $z<z_{\rm LS}$. For the purpose of the present work, we will assume that the matter and radiation densities at the CMB time correspond to those measured by the Planck collaboration. We will denote by $\overline H_0$ the value of $H_0$ that is inferred from the CMB data based on GR-$\Lambda$CDM cosmology. This value will generally differ from the value $H_0$ predicted by the models that deviate from GR-$\Lambda$CDM that are proposed in the following section. 

To simplify the analysis of the equations, it is convenient to express various contributions to the Friedman equation in units of $\overline H_0^2$. Consequently, for models deviating from GR-$\Lambda$CDM we define the following dimensionless quantities
\be\label{dide}
\mathring\rho_m (z)=\frac{G\rho_m(z)}{\overline H_0^2},\ \ \ \mathring\rho_r (z)=\frac{G\rho_r(z)}{\overline H_0^2}, 
\ee 
and
\be\label{lambdy}
\mathring\Lambda (z)=\frac{\Lambda(z)}{\overline H_0^2}.
\ee
The Friedman equation in the modified models becomes
\be\label{NewFriedman}
H(z)^2 = \overline H_0^2 \left[ \frac{8\pi }{3} \Big(\mathring\rho_m(z)+\mathring\rho_r(z)\Big)+ \frac{\mathring\Lambda(z)}{3} \right].
\ee
The evaluation of the Friedman equation at $z=0$ produces the value of the Hubble constant in the modified model $H_0^{}$, according to
\be \label{posta}
H_0^{}=\overline H_0 \sqrt{\frac{8\pi }{3} \Big(\mathring\rho_m(0)+\mathring\rho_r(0)\Big)+ \frac{\mathring\Lambda(0)}{3}}~.
\ee 

As noted above, in general, $H_0$ differs from $\overline H_0$ however, the direct substitution of \eqref{dide} and \eqref{lambdy} in \eqref{NewFriedman} clearly shows that the standard Friedman equation continues to hold for all values of $z$ and, in particular, at $ z=0$. 

In the following section, we will present some illustrative models that modify the GR-$\Lambda$CDM dynamics and exhibit the potential of these ideas to resolve the $ H_0$ tension.

\section{Two simple models} 
\label{s.models}

In this section we consider two simple models that, for a suitable choice of parameters, can resolve the $H_0$ tension. These models are simple illustrations of the potential interest of the approach presented here and are not meant to provide concrete, definite models but rather a framework to study the effects of diffusion in cosmology.  The general idea illustrated by these models paves the way for potentially multiple concrete realizations of specific forms of the diffusion mechanism which, in a realistic scenario, should be validated by further analysis.

More precisely, a convincing resolution of the $H_0$ tension will require on the one hand a clear understanding of the fundamental physics behind the diffusion process, together with a solid modelling of the cosmological and astrophysical conditions that produce it (possible steps aiming at such a scenario are proposed in \cite{Perez:2019gyd}), and on the other hand a Bayesian analysis comparing this framework to the standard GR-$\Lambda$CDM cosmology based on CMB, supernovae and baryon acoustic oscillation observations (note that this will require extending cosmological perturbation theory to include diffusion effects).  As the goal of this paper is to illustrate the potential viability of diffusion as a physical effect that could address the $H_0$ tension, and not to promote a specific diffusive mechanism, these two steps lie outside the scope of this paper and are left for future work.

Here we will consider two simple phenomenological models: one where the cosmological constant undergoes a sudden increase approximated by a step function, and another where the matter density decreases---due to energy transfer into the dark energy sector---at a faster rate than in GR-$\Lambda$CDM, and show that diffusion processes can offer a viable pathway to resolve the $H_0$ tension.

\begin{figure}[t]
\includegraphics[width=\columnwidth]{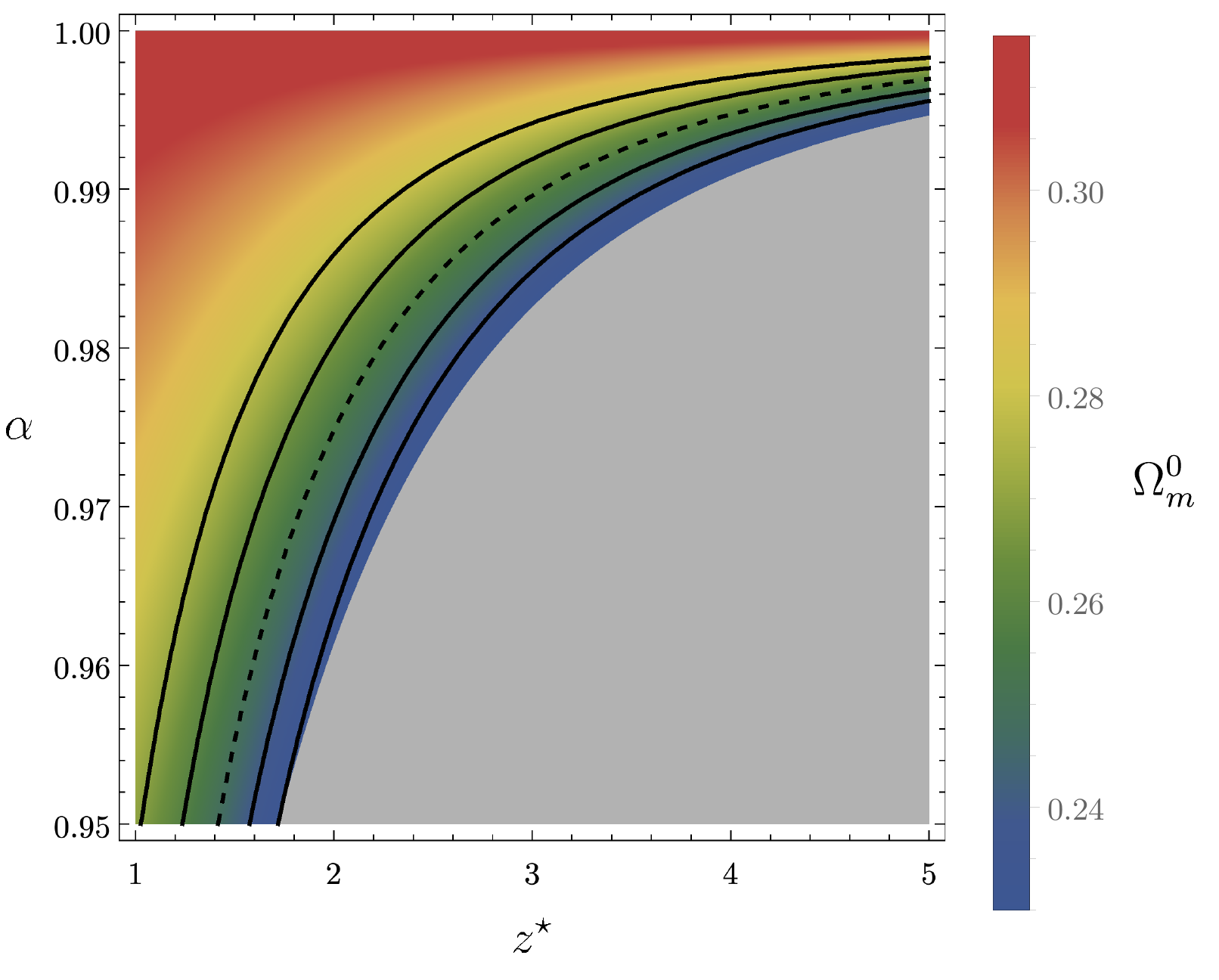}
\caption{This figure shows the region of the parameter space for $1 < z^\star < 5$ where the $H_0$ tension is alleviated or resolved for the sudden jump model with $\mathring\Lambda_\infty = 2.055$. Here $\alpha$ is plotted using a linear scale. The lines denote the 1$\sigma$ and 2$\sigma$ contours (assuming Gaussian uncertainty), and the dashed line indicates where the value of $H_0$ according to the sudden jump model exactly agrees with $H_0^{\rm sn}$. The predicted value of $\Omega_m^0$ according to the sudden jump model is shown by the colour (in the grey region, $\Omega_m^0 < 0.23$). Near $z^\star \sim 1.5$, in the best fit region $\Omega_m^0 \sim 0.25$; while near $z^\star \sim 5$, in the best fit region $\Omega_m^0 \sim 0.26$.}
\label{S0}
\end{figure}

\begin{figure}[t]
\includegraphics[width=\columnwidth]{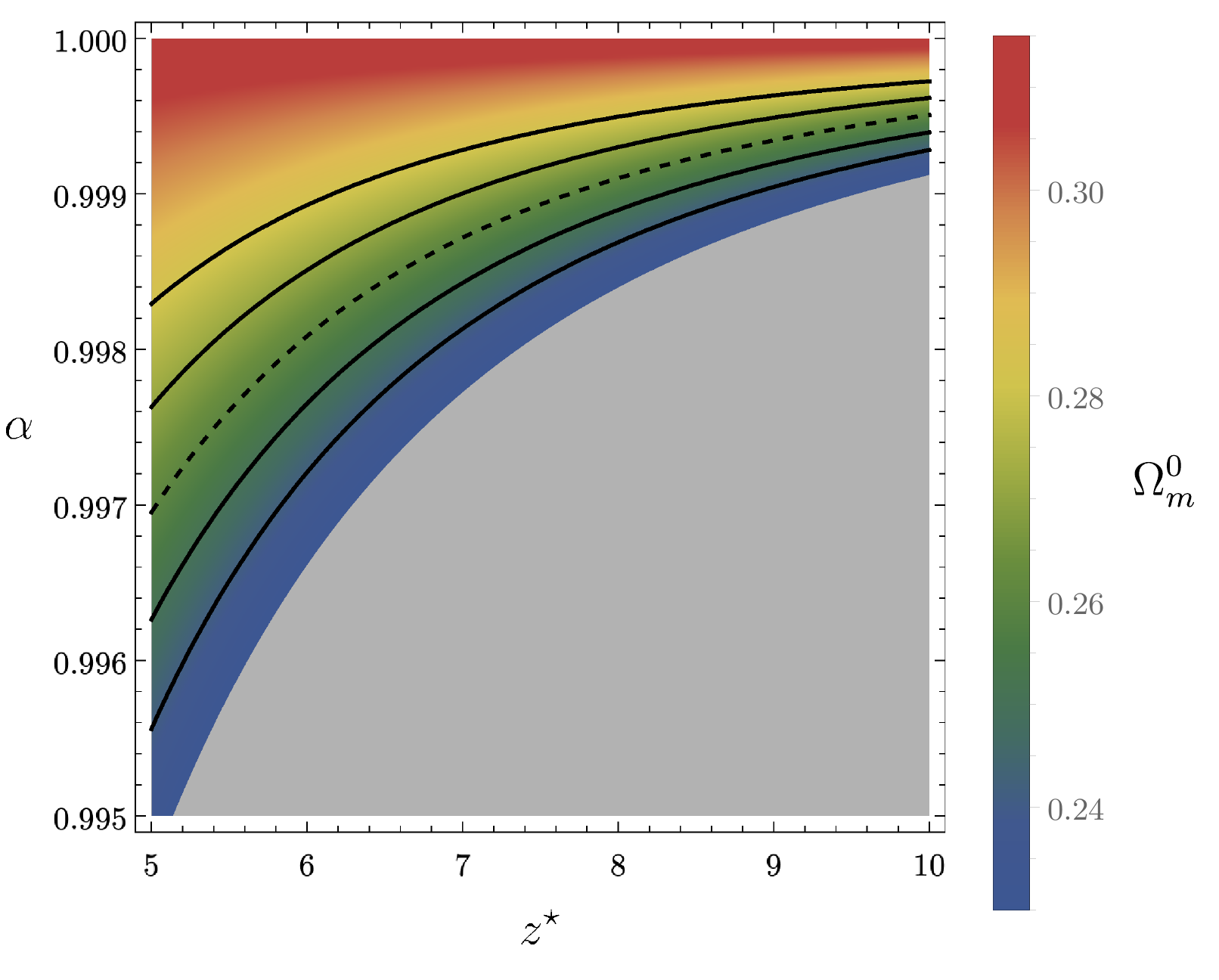}
\caption{This figure shows the region of the parameter space for $5 < z^\star < 10$ where the $H_0$ tension is alleviated or resolved for the sudden jump model with $\mathring\Lambda_\infty = 2.055$. Here $\alpha$ is plotted using a linear scale. The lines denote the 1$\sigma$ and 2$\sigma$ contours (assuming Gaussian uncertainty), and the dashed line indicates where the value of $H_0$ according to the sudden jump model exactly agrees with $H_0^{\rm sn}$. The predicted value of $\Omega_m^0$ according to the sudden jump model is shown by the colour (in the grey region, $\Omega_m^0 < 0.23$). In the best fit region $\Omega_m^0 \sim 0.26$.}
\label{S1}
\end{figure}

\begin{figure}[t]
\includegraphics[width=\columnwidth]{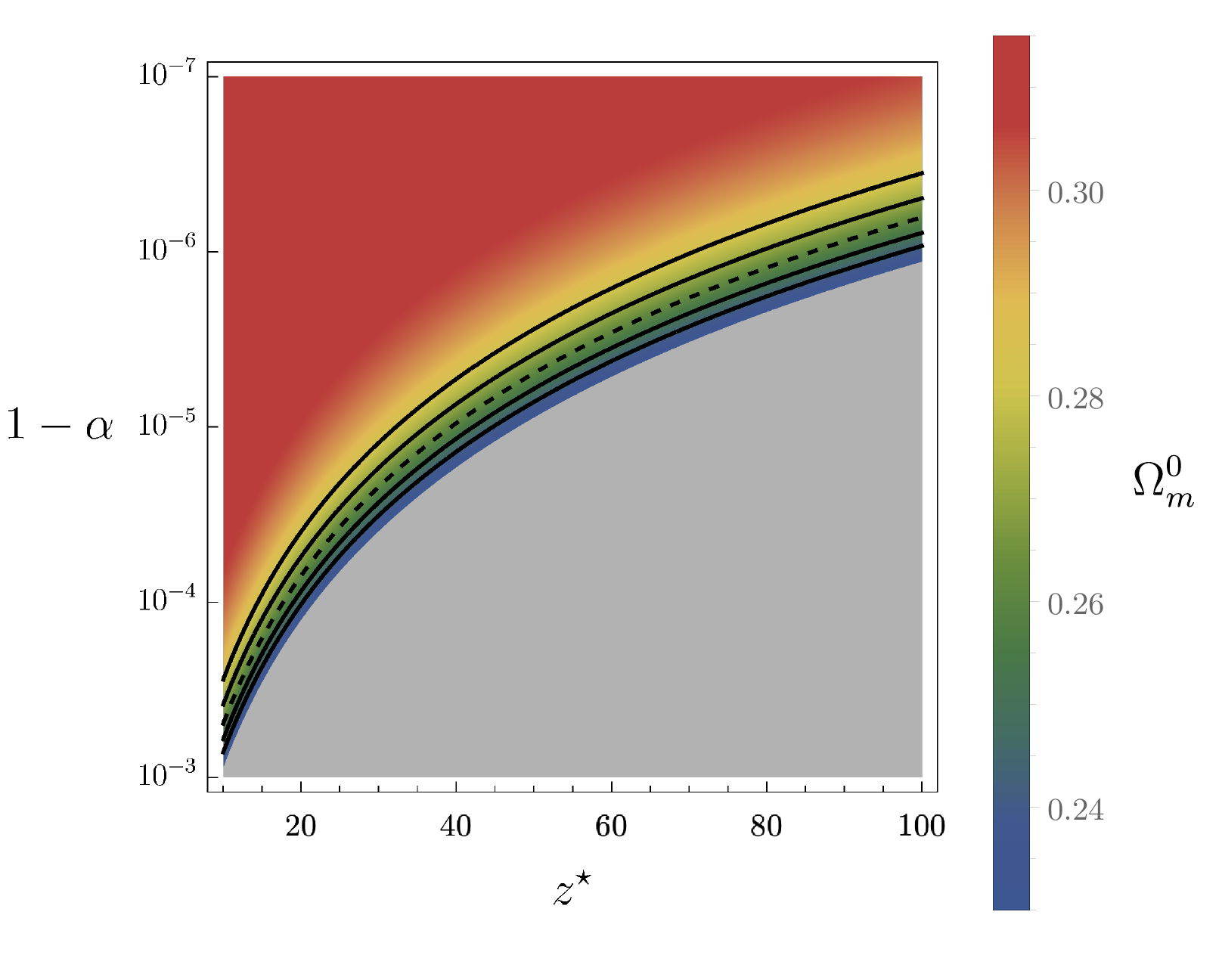}
\caption{This is the same diagram as in Figure \ref{S1}, but now considering larger values of $z^\star$ in the sudden transfer model and using a logarithmic scale for $1-\alpha$. As in Figure \ref{S1}, $\mathring\Lambda_\infty = 2.055$. In the best fit region $\Omega_m^0 \sim 0.26$.}
\label{S2}
\end{figure}

\begin{figure}[t]
\includegraphics[width=\columnwidth]{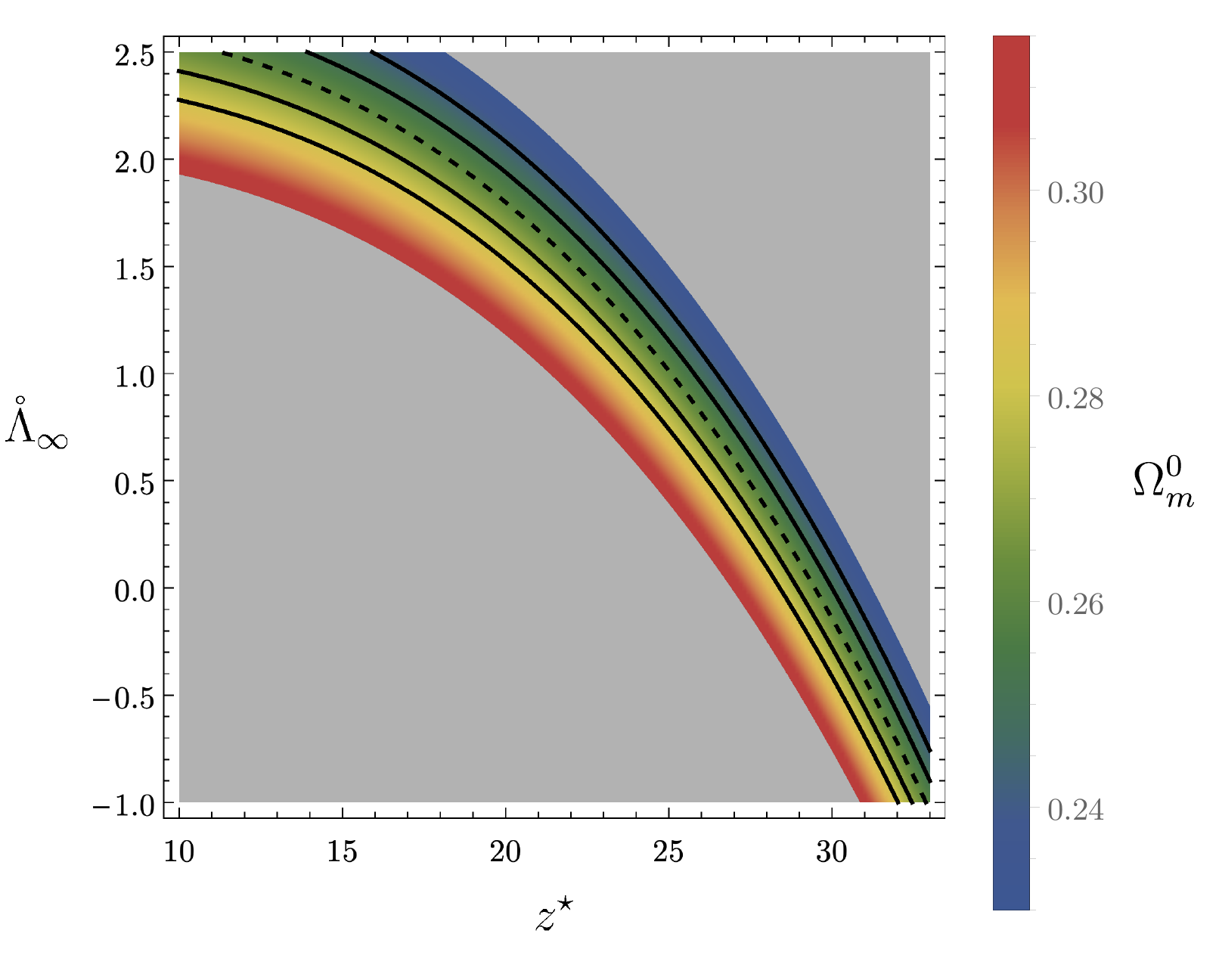}
\caption{This figure shows the region of parameter space where the $H_0$ tension is alleviated or resolved for the sudden jump model with $\alpha = 0.999$. Here $\mathring\Lambda_\infty$ is plotted using a linear scale. The lines denote the 1$\sigma$ and 2$\sigma$ contours (assuming Gaussian uncertainty), and the dashed line indicates where the value of $H_0$ according to the sudden jump model exactly agrees with $H_0^{\rm sn}$. The predicted value of $\Omega_m^0$ according to the sudden jump model is shown by the colour (in the grey region, $\Omega_m^0 < 0.23$ or $\Omega_m^0 > 0.315$).  Note that for $\alpha=0.999$ a negative value of $\mathring\Lambda_\infty$ is preferred for $z^\star \gtrsim 30$. In the best fit region $\Omega_m^0 \sim 0.26$.}
\label{S3}
\end{figure}

\subsection{Sudden transfer model}

A particularly simple model that captures the potential of the general framework to solve the $H_0$ tension is a sudden transfer of energy from the matter density to $\Lambda$ at some instant $z^\star$. Specifically,
\be
\rho_m(z) = \overline \rho_m^0 \, (1+z)^3 \, \Big[ \theta_+(z-z^\star) + \alpha \, \theta_-(z-z^\star) \Big],
\ee
with $0 < \alpha < 1$, and $\theta_{+}(x) $ is the Heaviside function (equal to unity for $x\ge 0$ and zero otherwise), and $\theta_-(x)=1-\theta_{+}(x)$. Here $\overline \rho_m^0$ is the energy density of matter today as inferred from the CMB data for a GR-$\Lambda$CDM cosmology. Note that due to the form of $\rho_m(z)$, this sudden jump model predicts the same value of $\rho_m$ at the surface of last scattering as in GR-$\Lambda$CDM.

This model represents a simple idealization of a rapid diffusion process (instantaneous in terms of the scale of the universe but that could be extended in astrophysical time depending on the value of $z^\star$).  Sudden transition models where the energy density of dark matter increases very rapidly have also been considered in, e.g., \cite{Bruni:2012sn}.

To simplify calculations and comparisons with the GR-$\Lambda$CDM model, it is convenient to work in terms of $\mathring \rho_m(z)$ defined in \eqref{dide},
\ba\label{bali}
\mathring \rho_m(z) &\equiv& \frac{G \rho_m(z)}{\overline H^2_0} \\ \n
&=& \frac{3\overline\Omega_m^0}{8\pi} (1+z)^3  \Big[\theta_+(z-z^\star) + \alpha \, \theta_-(z-z^\star) \Big].
\ea
Solving the modified continuity equation \eqref{difu} gives
\be \label{lali}
\mathring \Lambda(z)\equiv\frac{\Lambda(z)}{\overline H^2_0}= \mathring\Lambda_\infty + \mathring\Delta \, \theta_-(z-z^\star) ,
\ee
with
\be
\mathring\Delta = 3 (1-\alpha) (1+z^\star)^3 \,\, \overline \Omega_m^0.
\ee
The standard GR-$\Lambda$CDM cosmology is recovered for $\alpha=1$, and in this limit $\mathring\Lambda_\infty = 2.055$ gives $\Omega_\Lambda^0 = 0.685$.

Using equations \eqref{bali} and \eqref{lali}, $H_0$ can be computed from \eqref{posta}. Due to the sudden drop in the matter energy density at $z^\star$, the contribution of matter to the Friedman equation today as predicted by this model will be smaller than the value predicted by GR-$\Lambda$CDM, which implies that $\alpha<1$ for the ratio
\be \label{sj-alpha}
\alpha \equiv\frac{\rho^0_m}{\overline\rho_m^0}= \frac{\Omega_m^0 H_0^2}{\overline\Omega_m^0 \overline H_0^2} \
< 1.
\ee
Note that the value of $\Lambda$ for $z>z^\star$ could be positive or negative, depending on the value of $\mathring\Lambda_\infty$. In the results plotted in Figs.~\ref{S0}--\ref{S2}, we consider a positive $\mathring\Lambda_\infty$, while in Fig.~\ref{S3} a negative $\mathring\Lambda_\infty$ is allowed and is in fact preferred for larger $z^\star$.  Note that in the case of a negative $\mathring\Lambda_\infty$, as the cosmological constant was initially negative, there would occur a dynamical transition from an early 
negative $\Lambda$ era to the current (late-time) nearly de Sitter space-time, as proposed in \cite{Akarsu:2019hmw}.

To find which parameters for the sudden transfer model resolve the $H_0$ tension without requiring any modifications to pre-CMB physics, we demand that the inferred result for $H_0$---computed from \eqref{posta}---be compatible with $H_0^{\rm sn}$ (the value directly measured from local supernovae data). The results are shown in Figs.~\ref{S0}--\ref{S3}. In Figs.~\ref{S0}--\ref{S2}, the best fit region for $\mathring\Lambda_\infty = 2.055$ is shown (for different values of $z^\star$), while in Fig.~\ref{S3}, the best fit region for $\alpha = 0.999$ is shown. In Figs.~\ref{S1}--\ref{S3}, the best fit region corresponding to a predicted value of $H_0$ in agreement with supernova observations implies $\Omega_m^0 \sim 0.26$ for the sudden jump model, and a $\sim1\sigma$ departure occurs for $\Omega_m^0 \sim 0.27$. However, this is no longer the case if smaller values of $z^\star$ of order 1 are considered in which case lower values of $\Omega_m^0$ are preferred as can be seen in Fig.~\ref{S0}; for example, for $z^\star \sim 1$ the best fit region has $\Omega_m^0 \sim 0.24$ and for $z^\star \sim 1.5$ the best fit region has $\Omega_m^0 \sim 0.25$. Also, note that for fixed $\mathring\Lambda_\infty$, as $z^\star$ becomes larger $\alpha$ becomes closer and closer to 1 in the region where the $H_0$ tension is resolved.

\subsection{Anomalous decay of the matter density}

For the second phenomenological model, we assume that 
\be
\!\!\! \rho_m(z)=\overline\rho_m(z) \Bigg[ \theta_+(z-z^\star) + \left(\frac{ 1+z}{1+z^\star}\right)^{\!\! \gamma} \theta_-(z-z^\star) \Bigg],
\ee
with $\overline\rho_m(z) = \overline\rho_m^0  \, (1+z)^3$. In this model, the matter density in the past of $z^\star$ behaves like normal dust (with an initial value on the surface of last scattering matching the predicted value in the GR-$\Lambda$CDM model), but from $z^\star$ on (to smaller values of $z$) diffusion decreases the matter density anomalously as parametrized by $\gamma$. The energy lost is captured by dark energy according to equation \eqref{difu}. 

In terms of the dimensionless density \eqref{dide},
\ba
\! \mathring \rho_m(z) &&\equiv \frac{G\rho_m(z)}{{\overline H_0^2}} \\ \n
=&& \, \frac{3\overline \Omega_m^0}{8\pi} (1+z)^3 \Bigg[ \theta_+(z-z^\star) +
\bigg(\frac{1+z}{1+z^\star}\bigg)^{\!\!\gamma} \theta_-(z-z^\star) \Bigg],
\ea
where, because this model is based on a deformation of the standard $\Lambda$CDM model, the $\Lambda$CDM matter contribution to the critical density parameter $\overline\Omega_m^0$ in \eqref{lilo} naturally appears. As matter is diffusing into the dark energy sector,
\be\label{rrr}
\alpha \equiv\frac{\rho^0_m}{\overline\rho_m^0}= \frac{\Omega_m^0 H_0^2}{\overline\Omega_m^0 \overline H_0^2} =(1+z^\star)^{-\gamma}\ < 1,
\ee 
where the ratio $\alpha$ is expressed as a function of the parameters of the new model in the last equality.

\begin{figure}[t]
\includegraphics[width=\columnwidth]{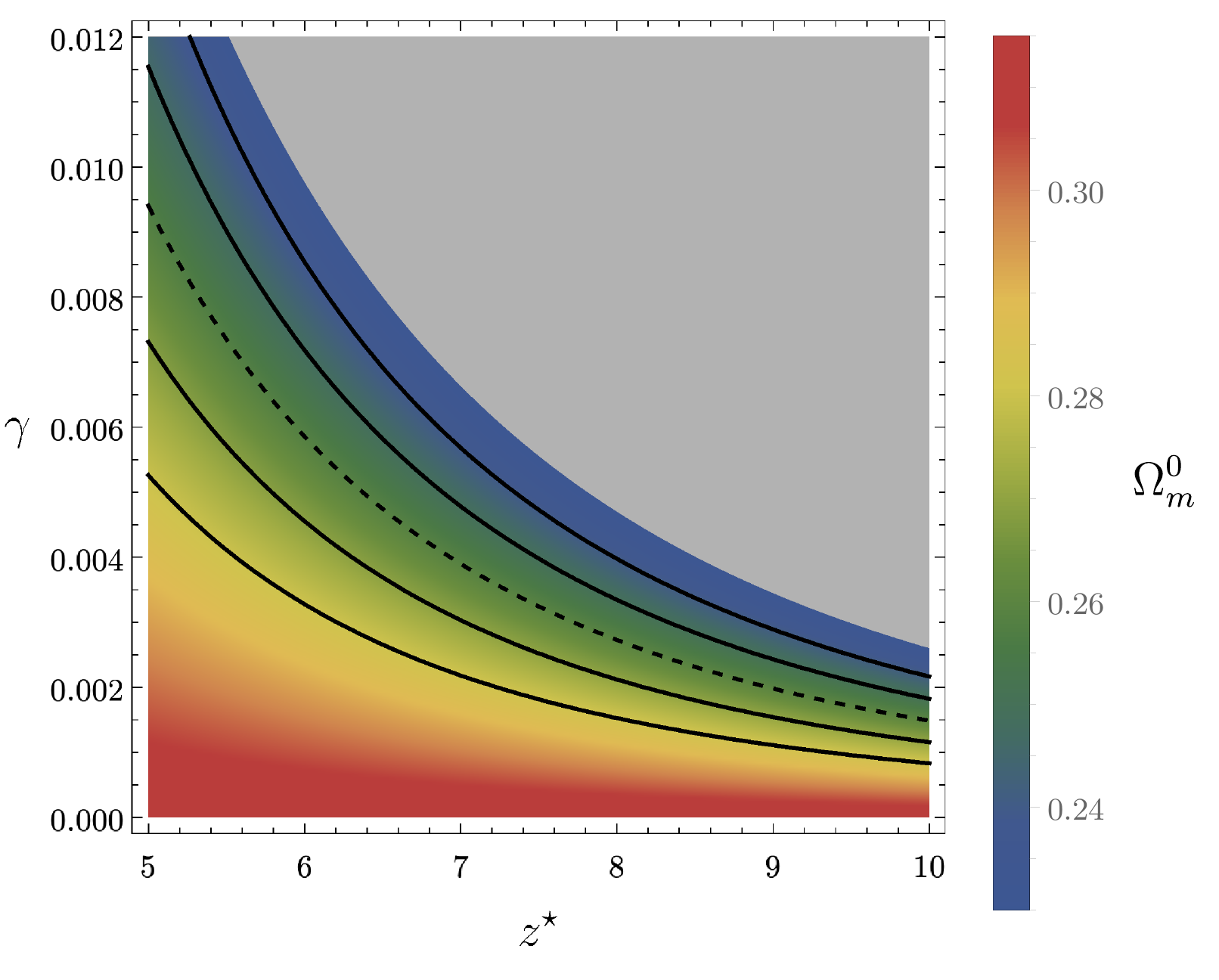}
\caption{This figure shows the region of the parameter space for small $z^\star$ where the $H_0$ tension is alleviated or resolved for the anomalous decay model with $\mathring\Lambda_\infty = 2.055$. Here $\gamma$ is plotted using a linear scale. The lines denote the 1$\sigma$ and 2$\sigma$ contours (assuming Gaussian uncertainty), and the dashed line indicates where the value of $H_0$ according to the sudden jump model exactly agrees with $H_0^{\rm sn}$. The predicted value of $\Omega_m^0$ according to the sudden jump model is shown by the colour (in the grey region, $\Omega_m^0 < 0.23$). In the best fit region $\Omega_m^0 \sim 0.26$.}
\label{M1}
\end{figure}

The solution to the continuity equation then implies that, in terms of the dimensionless quantity \eqref{lambdy},
\ba
\mathring \Lambda(z) &&\equiv \frac{\Lambda(z)}{{\overline H_0^2}} \\ \n =&& {\mathring\Lambda_{\infty}} - {\frac{3 \gamma \, {\overline\Omega_m^0}}{(\gamma+3)}\left[ \left(\frac{z+1}{z^\star+1}\right)^\gamma (z+1)^3-(z^\star+1)^3\right]},
\ea
for $z \le z^{\star}$ and
\be
\mathring\Lambda(z)={\mathring\Lambda_{\infty}},
\ee
for $z \ge z^\star$. As before, $\mathring\Lambda_\infty$ is a constant of integration and a free parameter in the model. The standard GR-$\Lambda$CDM cosmology is recovered for $\gamma=0$, and in this limit $\mathring\Lambda_\infty = 2.055$ gives $\Omega_\Lambda^0 = 0.685$.

\begin{figure}[t]
\includegraphics[width=\columnwidth]{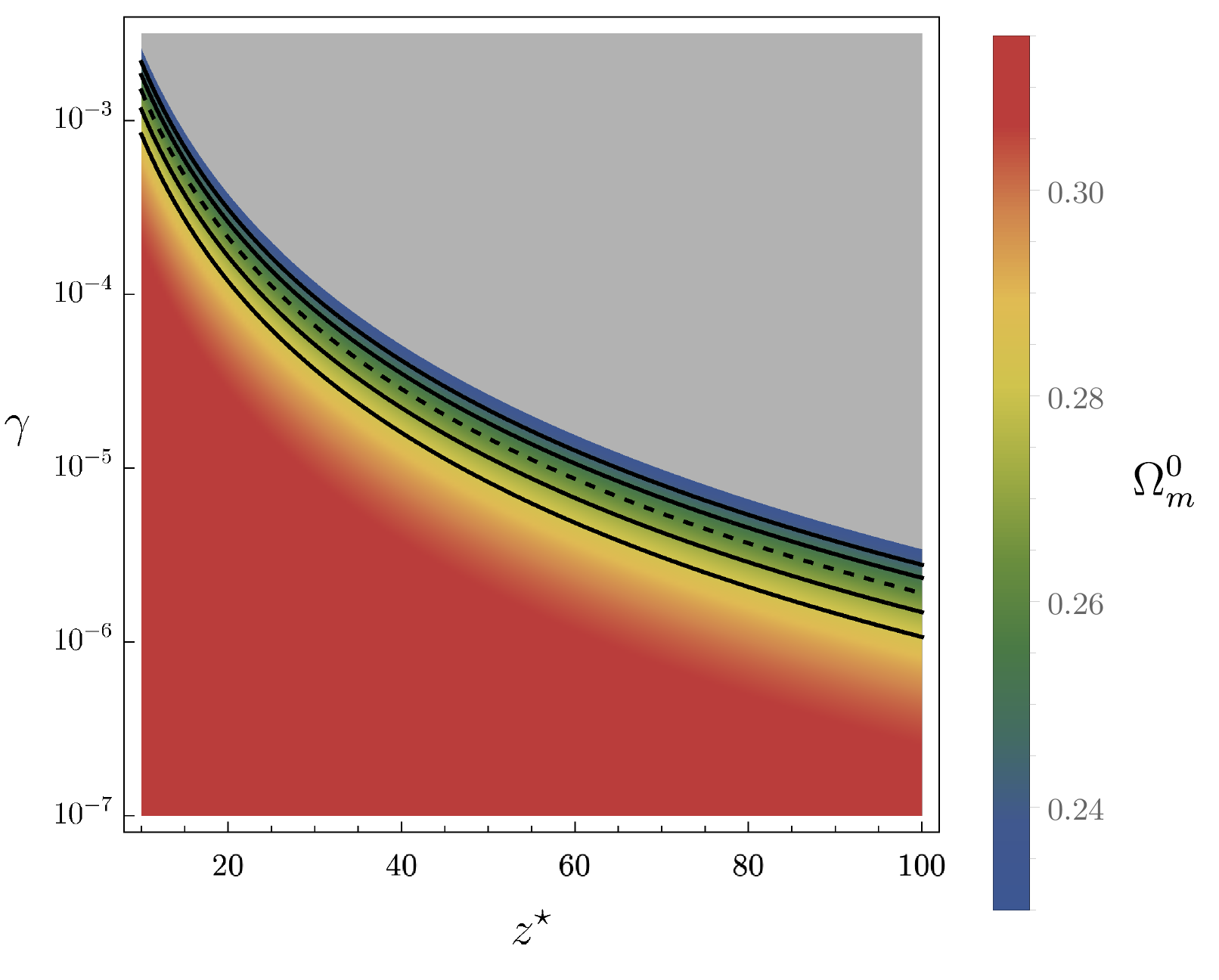}
\caption{This is the same diagram as in Figure \ref{M1}, but now considering larger values of $z^\star$ in the anomalous decay model and using a logarithmic scale for $\gamma$. As in Figure \ref{M1}, $\mathring\Lambda_\infty = 2.055$. As for smaller values of $z^\star$, in the best fit region $\Omega_m^0 \sim 0.26$.}
\label{M2}
\end{figure}

\begin{figure}[t]
\includegraphics[width=\columnwidth]{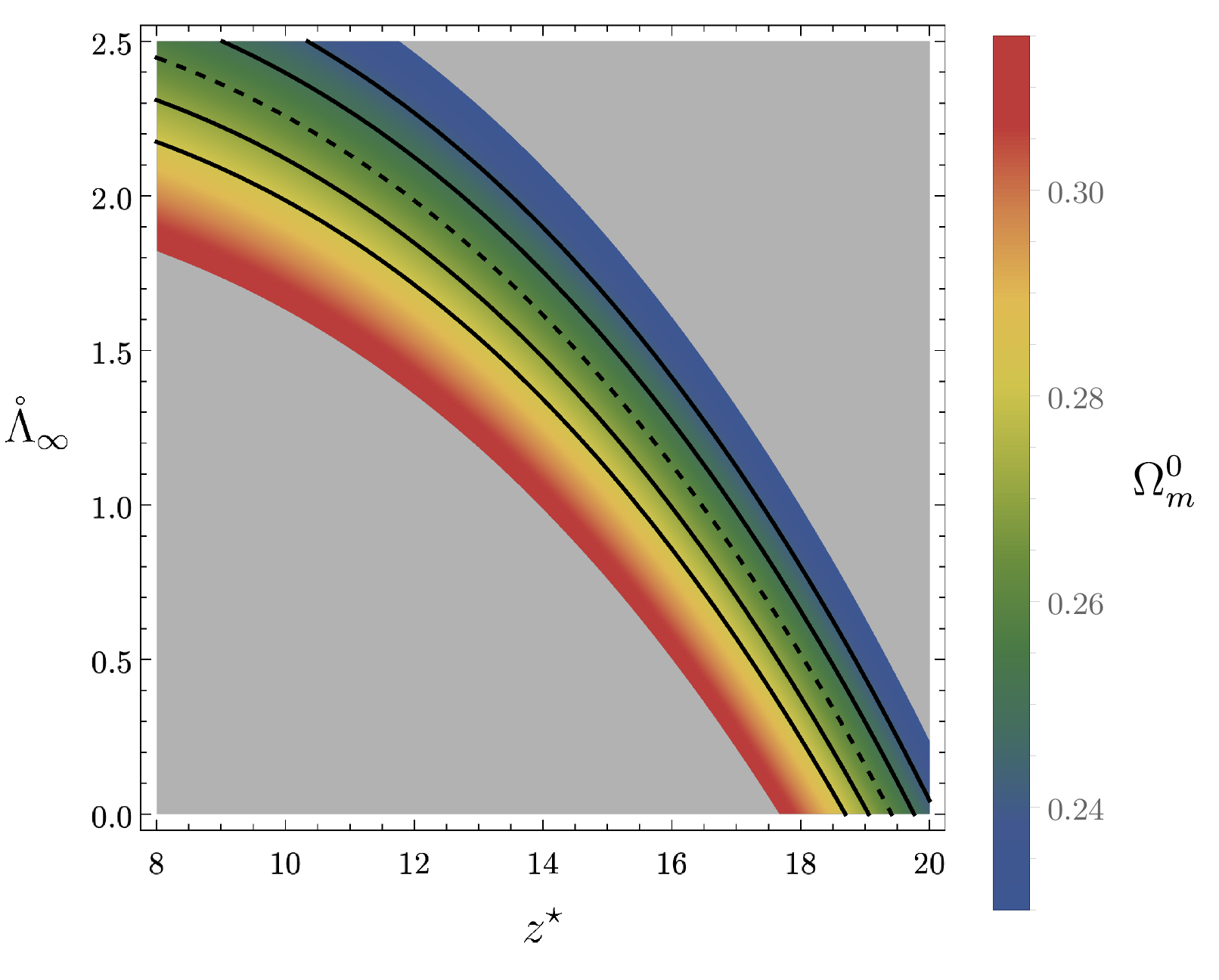}
\caption{This figure shows the region of parameter space where the $H_0$ tension is alleviated or resolved for the anomalous decay model with $\gamma = 10^{-3}$. Here $\mathring\Lambda_\infty$ is plotted using a linear scale. The lines denote the 1$\sigma$ and 2$\sigma$ contours (assuming Gaussian uncertainty), and the dashed line indicates where the value of $H_0$ according to the sudden jump model exactly agrees with $H_0^{\rm sn}$. The predicted value of $\Omega_m^0$ according to the sudden jump model is shown by the colour (in the grey region, $\Omega_m^0 < 0.23$ or $\Omega_m^0 > 0.315$). In the best fit region $\Omega_m^0 \sim 0.26$.}
\label{M3}
\end{figure}

As for the first model, to find the parameters for which the anomalous decay model can resolve the $H_0$ tension without requiring any modifications to pre-CMB physics, we calculate the inferred value for $H_0$ using \eqref{posta} and require that it be compatible with $H_0^{\rm sn}$; the results are shown in Figs.~\ref{M1}--\ref{M3}. In Figs.~\ref{M1} and \ref{M2}, the best fit region for $\mathring\Lambda_\infty = 2.055$ is shown (for small and large values of $z^\star$ respectively), while in Fig.~\ref{M3}, the best fit region for $\gamma = 10^{-3}$ is shown. Note that for a fixed choice of $\mathring\Lambda_\infty$, as $z^\star$ increases a smaller value of $\gamma$ preferred.

In all three figures, the best fit region corresponding to a predicted value of $H_0$ in agreement with supernova observations implies $\Omega_m^0 \sim 0.26$ and a $\sim1\sigma$ departure occurs for $\Omega_m^0 \sim 0.27$; this is similar to what was found for the sudden jump model. Again, this is no longer the case for smaller $z^\star \sim 1$ in which case the best fit region has a smaller $\Omega_m^0$.

Note that it is not surprising that (for $z^\star \gtrsim 5$) the best fit regions for $H_0$ in both of the two models considered here give $\Omega_m^0 \sim 0.26$. This is because in both cases $\alpha \approx 1$ for the best fit region (see Fig.~\ref{S2} for the sudden transfer model and Fig.~\ref{M2} together with Eq.~\eqref{rrr} for the anomalous decay model) and therefore $\Omega_m^0 \approx (\overline H_0^2/H_0^2) \overline\Omega_m^0 \approx 0.26$ for $H_0$ close to the value preferred by supernovae observations. Note that for smaller values of $z^\star$, the best fit region will have a smaller value for $\alpha$ (see for example Fig.~\ref{S0}), and therefore a smaller predicted $\Omega_m^0$ as well.

\section{Observations at intermediate redshifts}
\label{s.int-z}

It may be possible to have a `direct' detection of a growing $\Lambda$ from observations at intermediate redshifts $1 < z < 10$.

From the Friedman and Raychaudhuri equations for unimodular gravity with a varying $\Lambda$ \eqref{ug-fr}, and considering that pressureless matter together with $\Lambda$ dominate the dynamics (i.e., $\rho = \rho_m$ and $p=0$), the deceleration and jerk parameters defined in \eqref{taylor-a} would be given by
\be
q_0 = -1 + \f{3}{2} \Omega_m^0, \qquad
j_0 = 1 + \f{\dot \Lambda(0)}{2 H_0^3}.
\ee
For $z>1$, it is better to do a Taylor expansion for observables like $d_L$ around $y = z/(1+z)$ rather than $z$ \cite{Cattoen:2007sk} (see also, e.g., \cite{Dunsby:2015ers} for other possible perturbative expansions); but in any case the prefactors to $y^n$ in this Taylor series depend only on $q_0, j_0$, etc.

It is clear that, in the limit where $\Lambda$ is constant, the standard GR-$\Lambda$CDM cosmological model is recovered with the standard prediction $j_0 = 1$. On the other hand, in the context of unimodular gravity, with energy being `transferred' from matter fields to the dark energy sector, we would have $\dot\Lambda > 0$, and thus $j_0 > 1$. We note that the effect of dissipation in unimodular gravity on the jerk parameter has also been studied in \cite{Garcia-Aspeitia:2019yni}.

Type Ia supernovae have been observed up to redshifts $z \approx 2.3$, and these observations strongly disfavour models with $q_0 \approx -0.55$ but all other terms in the expansion (including $j_0$) vanishing \cite{Riess:2017lxs}. Future observations may serve to solidly constrain the value of $j_0$.

Also, the papers \cite{Camarena:2019moy, Camarena:2019rmj} report a measurement of the deceleration parameter using local observations (supernovae of type Ia with redshifts $0.023\le z\le 0.15$) with
\be
q_0=-1.08 \pm 0.29,
\ee
which is $1.9 \sigma$ away from the $\Lambda$CDM value $\overline q_0=-0.55$. Taken at face value, this result would suggest $\Omega_m^0< 0.14$ (see also \cite{Colgain:2019pck}). Even though these results must be taken with caution, they illustrate the point that there is, at present, a rather important degree of uncertainty concerning the amount of (dark and baryonic) matter present in the universe today, a fact that seems to leave room for a level of energy diffusion of the magnitude required in the models considered here that can resolve the $H_0$ tension.

It is clear that observations at larger $z$ could potentially provide stronger constraints on the jerk parameter and potentially differentiate between the predictions of the GR-$\Lambda$CDM cosmology and UG models with diffusion. It must however be noted that such constraints can only be considered as valid within a particular functional form of the parametrizations appearing in these models, as it is clear that the value of the derivative of a function $f(z)$ at a certain point (say $z=0$) is compatible with infinitely many different functions.

Interestingly, it has recently been claimed that quasars and gamma ray bursts can also be used as standard candles (using a non-linear relation between ultraviolet and x-ray emission for quasars, and the Amati relation for gamma ray bursts), and the result of an analysis {of observations at redshifts $1 \lesssim z \lesssim 7$} gives $j_0 > 1$ at a 4$\sigma$ confidence level \cite{Lusso:2019akb} (however see also \cite{Yang:2019vgk, Velten:2019vwo}).

\newpage

\section{Discussion}
\label{s.disc}

We have presented a general scenario that can resolve the $H_0$ tension. The new perspective is motivated by the proposal in \cite{Perez:2018wlo, Perez:2017krv} for a fundamental account of the nature and value of dark energy. The model does not require problematic equations of state for the components of the cosmic pie. However, it does require violations of the local conservation of energy momentum (i.e., deviations from $\nabla^a T_{ab} = 0$), and thus a modified version of GR that can accommodate these: unimodular gravity. We have provided only a brief motivation for this possibility here, because that has been (and will be) the subject of other works. The present analysis has focused on the essential aspects of the phenomenology. The basic effect can be seen as an effective `flow of energy' from the matter sector to dark energy (or `cosmological constant') sector. The viability of the model depends on its ability to sucessfully overcome several constraints. It must not disrupt the conditions in the CMB in any substantial manner, avoiding any negative repercussion regarding the success of the GR-$\Lambda$CDM model. It must, of course, resolve or substantially reduce the $H_0$ tension, thus accommodating the late time supernovae observations, and it must not lead to an excessive reduction in amounts of the energy density of the matter sector. The last point is essential so as not to be in conflict with late time observations of that sector. In other words, the amount of dark matter that, according to the viable scenarios, was converted into dark energy in the period between the surface of last scattering and the present, cannot be more than a small fraction of its initial value.

Concerning this last point, CMB observations, combined with the assumption of GR-$\Lambda$CDM, imply that $\Omega^0_{\rm Baryons}\sim 0.05$ and $\Omega^0_{\rm DM} \sim 0.26$; however, direct present day observations indicate that most of the implied baryonic matter must be invisible (i.e., is not accounted for in the star gas and dust that forms the galaxies and the intergalactic media), namely $\Omega_{\rm Visible-Baryons}^0\sim 0.02$ \cite{Fukugita:1997bi}. (Some recent reports indicate that $\Omega_{\rm Visible-Baryons}^0$ might account for a larger fraction of $\Omega^0_{\rm Baryons}$ than previously suspected \cite{Nicastro:2018eam}, but substantial uncertainties remain \cite{Bregman_2018}.) On the other hand, the dark matter in galaxies and clusters is known to be about 10 times the amount of luminous matter (which itself is a fraction of the baryonic matter as a substantial part of it seems to be in the intergalactic media). Thus the portion of the dark matter component that we can directly infer from galaxy surveys today is of the order of $\sim 0.20$. That is, we have strong confidence that at present there is at least $\Omega^0_{\rm Matter} \sim 0.22$. Taking into account the claims of a substantial presence of baryonic matter in the intergalactic media, we could say that a conservative constraint in this regard ought to be $\Omega^0_{\rm Matter} \sim 0.25$. The point is that given the uncertainties in the amount of matter directly observed or indirectly inferred to be present in today's universe there seems to be room for an important missing fraction compared to the GR-$\Lambda$CDM value of $\Omega^0_{\rm Matter} \sim 0.31$. Thus we consider that our models are viable in this regard if the fraction of $\Omega^0_{\rm Matter}$ that was lost via our mechanism in the period from the last scattering surface to the present is smaller than, say, $20\%$. As can be seen in the figures of the previous sections, the models we have presented pass this test for various ranges of the parameters compatible with the other requirements. Notice that the strong constraints on $\Omega_{\rm Baryons}^0$ coming from nucleosynthesis are not relevant in the present analysis because all the modifications introduced by our models happen after recombination, and diffusion effects can cause the amplitude of $\rho_{\rm Baryons}$ to decrease faster than what would be expected from cosmological expansion alone.

The simplicity of the two models considered here and their effectiveness in resolving the $H_0$ tension illustrates the potential of the perspective that we put forward here. A more realistic model would require, in addition, a detailed account of the fundamental mechanism behind the diffusion process as well as the astrophysical description of the sources, their abundance and dynamical evolution during the recent cosmological history. Thus, the framework presented here naturally calls for further studies involving the simultaneous analysis of parametrizations characterizing these two combined aspects of the problem, with best fit studies based on the wide range of available data regarding cosmological observations. One such possibility of more complete scenario is the natural generalization of the fundamental mechanism of \cite{Perez:2018wlo, Perez:2017krv} that would include black holes as diffusive sources \cite{Perez:2019gyd}. We should note that proposals having some resemblance to the present one, but motivated by rather different physical ideas have been put forward in \cite{Garcia-Aspeitia:2019yni, Garcia-Aspeitia:2019yod}.

More generally, unimodular gravity, combined with the relaxation of the assumption of conservation of the energy-momentum tensor of matter, represents, at a classical level, a relatively mild modification of GR which ought to be subjected to studies and constraining observational tests, just like other modified gravity theories. The present work together with the results of \cite{Perez:2018wlo, Perez:2017krv} clearly exhibits its potential.

Finally, note that in this paper we have been working under the assumption that the cosmological parameters at the last scattering surface are those determined by the Planck analysis of their data. That analysis in turn has been carried out under the assumption of the validity of the standard GR-$\Lambda$CDM model. Our modified model calls for a reevaluation of those parameters, given the Planck observations, but carried out with the modified cosmic evolution, in particular by using a Boltzmann code updated to include the diffusion effects of either of the models considered here (or perhaps another model entirely). This would naturally lead to small changes in the preferred values of some cosmological parameters, but we expect these resulting changes to be of higher order and lead only to small modifications of the results derived here based on the ratio of the sound horizon and the radius of the surface of last scattering. That more complete and complex analysis will be the subject of a forthcoming paper.

\acknowledgments

We thank Marco de Cesare and Violaine Ponsin for help with the figures, and E.W.-E.~thanks Aix-Marseille Universit\'e for hospitality during the initial stages of this work.
E.W.-E.~is supported in part by the Natural Science and Engineering Research Council of Canada and by a Harrison McCain Foundation Young Scholars Award.  
D.S.~acknowledges partial financial support from PAPIIT-UNAM (M\'exico) Grant No.~IG100120; CONACYT (M\'exico) project ``Frontiers of Science'' No.~140630; the Foundational Questions Institute (Grant No.~FQXi-MGB-1928); and the Fetzer Franklin Fund, a donor advised by the Silicon Valley Community Foundation.

\providecommand{\href}[2]{#2}\begingroup\raggedright\endgroup


\begin{thebibliography}{10}

\bibitem{Adam:2015rua}
{\bf Planck Collaboration}, R.~Adam {\em et al.}, ``{Planck 2015 results. I.
  Overview of products and scientific results},'' Astron. Astrophys. {\bf 594}
  (2016) A1,
\href{http://arXiv.org/abs/1502.01582}{{\tt arXiv:1502.01582}}.

\bibitem{Aghanim:2018eyx}
{\bf Planck Collaboration}, N.~Aghanim {\em et al.}, ``{Planck 2018 results.
  VI. Cosmological parameters},''
\href{http://arXiv.org/abs/1807.06209}{{\tt arXiv:1807.06209}}.

\bibitem{Alam:2016hwk}
{\bf BOSS Collaboration}, S.~Alam {\em et al.}, ``{The clustering of galaxies
  in the completed SDSS-III Baryon Oscillation Spectroscopic Survey:
  cosmological analysis of the DR12 galaxy sample},'' Mon. Not. Roy. Astron.
  Soc. {\bf 470} (2017) 2617--2652,
\href{http://arXiv.org/abs/1607.03155}{{\tt arXiv:1607.03155}}.

\bibitem{Riess:2019cxk}
A.~G. Riess, S.~Casertano, W.~Yuan, L.~M. Macri, and D.~Scolnic, ``{Large
  Magellanic Cloud Cepheid Standards Provide a 1\% Foundation for the
  Determination of the Hubble Constant and Stronger Evidence for Physics beyond
  $\Lambda$CDM},'' Astrophys. J. {\bf 876} (2019) 85,
\href{http://arXiv.org/abs/1903.07603}{{\tt arXiv:1903.07603}}.

\bibitem{Reid:2019tiq}
M.~Reid, D.~Pesce and A.~Riess, ``{An Improved Distance to NGC 4258 and its
  Implications for the Hubble Constant},'' Astrophys. J. \textbf{886} (2019) L27,
\href{http://arXiv.org/abs/1908.05625}{{\tt arXiv:1908.05625}}.

\bibitem{Freedman:2019jwv}
W.~L. Freedman {\em et al.}, ``{The Carnegie-Chicago Hubble Program. VIII. An
  Independent Determination of the Hubble Constant Based on the Tip of the Red
  Giant Branch},''
\href{http://arXiv.org/abs/1907.05922}{{\tt arXiv:1907.05922}}.

\bibitem{Farr:2019twy}
W.~M. Farr, M.~Fishbach, J.~Ye, and D.~Holz, ``A Future Percent-Level
  Measurement of the Hubble Expansion at Redshift 0.8 With Advanced LIGO,''
  Astrophys. J. {\bf 883} (2019) L42,
  \href{http://arXiv.org/abs/1908.09084}{{\tt arXiv:1908.09084}}.

\bibitem{Bergstrom:2000pn}
L.~Bergstr{\"o}m, ``Nonbaryonic dark matter: Observational evidence and
  detection methods,'' Rept. Prog. Phys. {\bf 63} (2000) 793,
  \href{http://arXiv.org/abs/hep-ph/0002126}{{\tt arXiv:hep-ph/0002126}}.

\bibitem{Clifton:2011jh}
T.~Clifton, P.~G. Ferreira, A.~Padilla, and C.~Skordis, ``Modified Gravity and
  Cosmology,'' Phys. Rept. {\bf 513} (2012) 1--189,
  \href{http://arXiv.org/abs/1106.2476}{{\tt arXiv:1106.2476}}.

\bibitem{Martin:2012bt}
J.~Martin, ``Everything You Always Wanted To Know About The Cosmological
  Constant Problem (But Were Afraid To Ask),'' Comptes Rendus Physique {\bf 13}
  (2012) 566--665, \href{http://arXiv.org/abs/1205.3365}{{\tt
  arXiv:1205.3365}}.

\bibitem{Riess:1998cb}
{\bf Supernova Search Team}, A.~G. Riess {\em et al.}, ``{Observational
  evidence from supernovae for an accelerating universe and a cosmological
  constant},'' Astron. J. {\bf 116} (1998) 1009--1038,
\href{http://arXiv.org/abs/astro-ph/9805201}{{\tt arXiv:astro-ph/9805201}}.

\bibitem{Perlmutter:1998np}
{\bf Supernova Cosmology Project}, S.~Perlmutter {\em et al.}, ``{Measurements
  of Omega and Lambda from 42 high redshift supernovae},'' Astrophys. J. {\bf
  517} (1999) 565--586,
\href{http://arXiv.org/abs/astro-ph/9812133}{{\tt arXiv:astro-ph/9812133}}.

\bibitem{Bernal:2016gxb}
J.~L. Bernal, L.~Verde, and A.~G. Riess, ``{The trouble with $H_0$},'' JCAP
  {\bf 1610} (2016) 019,
\href{http://arXiv.org/abs/1607.05617}{{\tt arXiv:1607.05617}}.

\bibitem{Freedman:2017yms}
W.~L. Freedman, ``{Cosmology at a Crossroads},'' Nat. Astron. {\bf 1} (2017)
  0121,
\href{http://arXiv.org/abs/1706.02739}{{\tt arXiv:1706.02739}}.

\bibitem{Poulin:2018cxd}
V.~Poulin, T.~L. Smith, T.~Karwal, and M.~Kamionkowski, ``{Early Dark Energy
  Can Resolve The Hubble Tension},'' Phys. Rev. Lett. {\bf 122} (2019) 221301,
\href{http://arXiv.org/abs/1811.04083}{{\tt arXiv:1811.04083}}.

\bibitem{Colgain:2018wgk}
E.~\'O Colg\'ain, M.~H.~van Putten and H.~Yavartanoo, ``{de Sitter Swampland, $H_0$
  tension \& observation},'' Phys.\ Lett.\ \textbf{B793} (2019) 126-129,
\href{http://arXiv.org/abs/1807.07451}{{\tt arXiv:1807.07451}}.

\bibitem{Kumar:2019wfs}
  S.~Kumar, R.~C.~Nunes and S.~K.~Yadav,
  ``{Dark sector interaction: a remedy of the tensions between CMB and LSS data},''
  Eur.\ Phys.\ J.\ {\bf C79} (2019) 576,
\href{http://arXiv.org/abs/1903.04865}{{\tt arXiv:1903.04865}}.

\bibitem{DiValentino:2019ffd}
E.~Di~Valentino, A.~Melchiorri, O.~Mena, and S.~Vagnozzi, ``{Interacting dark
  energy after the latest Planck, DES, and $H_0$ measurements: an excellent
  solution to the $H_0$ and cosmic shear tensions},''
\href{http://arXiv.org/abs/1908.04281}{{\tt arXiv:1908.04281}}.

\bibitem{Agrawal:2019dlm}
P.~Agrawal, G.~Obied, and C.~Vafa, ``{$H_0$ Tension, Swampland Conjectures and
  the Epoch of Fading Dark Matter},''
\href{http://arXiv.org/abs/1906.08261}{{\tt arXiv:1906.08261}}.

\bibitem{Rossi:2019lgt}
M.~Rossi, M.~Ballardini, M.~Braglia, F.~Finelli, D.~Paoletti, A.~A.~Starobinsky
  and C.~Umilt\`a, ``{Cosmological constraints on post-Newtonian parameters in
  effectively massless scalar-tensor theories of gravity},'' Phys.\ Rev.\
  {\bf D100} (2019) 103524,
\href{http://arXiv.org/abs/1906.10218}{{\tt arXiv:1906.10218}}.

\bibitem{Josset:2016vrq}
T.~Josset, A.~Perez, and D.~Sudarsky, ``{Dark Energy from Violation of Energy
  Conservation},'' Phys. Rev. Lett. {\bf 118} (2017) 021102,
\href{http://arXiv.org/abs/1604.04183}{{\tt arXiv:1604.04183}}.

\bibitem{Collins:2004bp}
J.~Collins, A.~Perez, D.~Sudarsky, L.~Urrutia, and H.~Vucetich, ``{Lorentz
  invariance and quantum gravity: an additional fine-tuning problem?},''
  Phys.Rev.Lett. {\bf 93} (2004) 191301,
\href{http://arXiv.org/abs/gr-qc/0403053}{{\tt arXiv:gr-qc/0403053}}.

\bibitem{Collins:2006bw}
J.~Collins, A.~Perez, and D.~Sudarsky, ``{Lorentz invariance violation and its
  role in quantum gravity phenomenology},''
\href{http://arXiv.org/abs/hep-th/0603002}{{\tt arXiv:hep-th/0603002}}.

\bibitem{Perez:2017krv}
A.~Perez and D.~Sudarsky, ``{Dark energy from quantum gravity discreteness},''
  Phys. Rev. Lett. {\bf 122} (2019) 221302,
\href{http://arXiv.org/abs/1711.05183}{{\tt arXiv:1711.05183}}.

\bibitem{Perez:2018wlo}
A.~Perez, D.~Sudarsky, and J.~D. Bjorken, ``{A microscopic model for an
  emergent cosmological constant},'' Int. J. Mod. Phys. {\bf D27} (2018)
  1846002,
\href{http://arXiv.org/abs/1804.07162}{{\tt arXiv:1804.07162}}.

\bibitem{Perez:2019gyd}
A.~Perez and D.~Sudarsky, ``{Black holes, Planckian granularity, and the
  changing cosmological 'constant'},''
\href{http://arXiv.org/abs/1911.06059}{{\tt arXiv:1911.06059}}.

\bibitem{Banerjee:2019kgu} 
S.~Banerjee, D.~Benisty and E.~I.~Guendelman, ``{Running Vacuum from Dynamical
  Spacetime Cosmology},''
\href{http://arXiv.org/abs/1910.03933}{{\tt arXiv:1910.03933}}.

\bibitem{Ellis:2010uc}
G.~F.~R. Ellis, H.~van Elst, J.~Murugan, and J.-P. Uzan, ``{On the Trace-Free
  Einstein Equations as a Viable Alternative to General Relativity},'' Class.
  Quant. Grav. {\bf 28} (2011) 225007,
\href{http://arXiv.org/abs/1008.1196}{{\tt arXiv:1008.1196}}.

\bibitem{Weinberg:1988cp}
S.~Weinberg, ``{The Cosmological Constant Problem},'' Rev. Mod. Phys. {\bf 61}
  (1989) 1--23.

\bibitem{Sorkin:2003bx}
R.~D.~Sorkin, ``{Causal sets: Discrete gravity},''
\href{http://arXiv.org/abs/gr-qc/0309009}{{\tt arXiv:gr-qc/0309009}}.

\bibitem{Jacobson:1995ab}
T.~Jacobson, ``{Thermodynamics of space-time: The Einstein equation of state},''
Phys. Rev. Lett. \textbf{75} (1995) 1260-1263,
\href{http://arXiv.org/abs/gr-qc/9504004}{{\tt arXiv:gr-qc/9504004}}.

\bibitem{Jacobson:2015hqa}
T.~Jacobson, ``{Entanglement Equilibrium and the Einstein Equation},''
Phys. Rev. Lett. \textbf{116} (2016) 201101,
\href{http://arXiv.org/abs/1505.04753}{{\tt arXiv:1505.04753}}.

\bibitem{Benisty:2017eqh} 
D.~Benisty and E.~I.~Guendelman, ``{Interacting Diffusive Unified Dark Energy and
  Dark Matter from Scalar Fields},'' Eur.\ Phys.\ J.\ {\bf C77} (2017) 396,
\href{http://arXiv.org/abs/1701.08667}{{\tt arXiv:1701.08667}}.

\bibitem{Benisty:2018oyy} 
D.~Benisty, E.~Guendelman and Z.~Haba, ``{Unification of dark energy and dark matter
  from diffusive cosmology},'' Phys.\ Rev.\ {\bf D99} (2019) 123521,
\href{http://arXiv.org/abs/1812.06151}{{\tt arXiv:1812.06151}}.

\bibitem{Eppley1977}
K.~Eppley and E.~Hannah, ``The necessity of quantizing the gravitational
  field,'' Found. Phys. {\bf 7} (1977) 51--68.

\bibitem{Page:1981aj}
D.~N. Page and C.~D. Geilker, ``{Indirect Evidence for Quantum Gravity},''
  Phys. Rev. Lett. {\bf 47} (1981)
979--982.

\bibitem{Carlip:2008zf}
S.~Carlip, ``{Is Quantum Gravity Necessary?},'' Class. Quant. Grav. {\bf 25}
  (2008) 154010,
\href{http://arXiv.org/abs/0803.3456}{{\tt arXiv:0803.3456}}.

\bibitem{Maudlin:2019bje}
T.~Maudlin, E.~Okon, and D.~Sudarsky, ``{On the Status of Conservation Laws in
  Physics: Implications for Semiclassical Gravity},''
\href{http://arXiv.org/abs/1910.06473}{{\tt arXiv:1910.06473}}.

\bibitem{Amadei:2019ssp}
L.~Amadei and A.~Perez, ``{Hawking's information puzzle: a solution realized in
  loop quantum cosmology},''
\href{http://arXiv.org/abs/1911.00306}{{\tt arXiv:1911.00306}}.

\bibitem{Amadei:2019wjp}
L.~Amadei, H.~Liu, and A.~Perez, ``{Unitarity and information in quantum
  gravity: a simple example},''
\href{http://arXiv.org/abs/1912.09750}{{\tt arXiv:1912.09750}}.

\bibitem{Perez:2014xca}
A.~Perez, ``{No firewalls in quantum gravity: the role of discreteness of
  quantum geometry in resolving the information loss paradox},'' Class. Quant.
  Grav. {\bf 32} (2015) 084001,
\href{http://arXiv.org/abs/1410.7062}{{\tt arXiv:1410.7062}}.

\bibitem{Perez:2017cmj}
A.~Perez, ``{Black Holes in Loop Quantum Gravity},'' Rept. Prog. Phys. {\bf 80}
  (2017) 126901,
\href{http://arXiv.org/abs/1703.09149}{{\tt arXiv:1703.09149}}.

\bibitem{Perez:2005gh}
A.~Perez, H.~Sahlmann, and D.~Sudarsky, ``{On the quantum origin of the seeds
  of cosmic structure},'' Class. Quant. Grav. {\bf 23} (2006) 2317--2354,
\href{http://arXiv.org/abs/gr-qc/0508100}{{\tt arXiv:gr-qc/0508100}}.

\bibitem{Leon:2017sru}
G.~Le\'on, ``{Eternal inflation and the quantum birth of cosmic structure},''
  Eur. Phys. J. {\bf C77} (2017) 705,
\href{http://arXiv.org/abs/1705.03958}{{\tt arXiv:1705.03958}}.

\bibitem{Okon:2013lsa}
E.~Okon and D.~Sudarsky, ``{Benefits of Objective Collapse Models for Cosmology
  and Quantum Gravity},'' Found. Phys. {\bf 44} (2014) 114--143,
\href{http://arXiv.org/abs/1309.1730}{{\tt arXiv:1309.1730}}.

\bibitem{Okon:2014dpa}
E.~Okon and D.~Sudarsky, ``{The Black Hole Information Paradox and the Collapse
  of the Wave Function},'' Found. Phys. {\bf 45} (2015) 461--470,
\href{http://arXiv.org/abs/1406.2011}{{\tt arXiv:1406.2011}}.

\bibitem{Modak:2014vya}
S.~K. Modak, L.~Ort\'iz, I.~Pe{\~n}a, and D.~Sudarsky, ``{Non-Paradoxical Loss of
  Information in Black Hole Evaporation in a Quantum Collapse Model},'' Phys.
  Rev. {\bf D91} (2015) 124009,
\href{http://arXiv.org/abs/1408.3062}{{\tt arXiv:1408.3062}}.

\bibitem{Bedingham:2016aus}
D.~Bedingham, S.~K. Modak, and D.~Sudarsky, ``{Relativistic collapse dynamics
  and black hole information loss},'' Phys. Rev. {\bf D94} (2016) 045009,
\href{http://arXiv.org/abs/1604.06537}{{\tt arXiv:1604.06537}}.

\bibitem{Okon:2017pvc}
E.~Okon and D.~Sudarsky, ``{Losing stuff down a black hole},'' Found. Phys.
  {\bf 48} (2018) 411--428,
\href{http://arXiv.org/abs/1710.01451}{{\tt arXiv:1710.01451}}.

\bibitem{Bassi_2005}
A.~Bassi, E.~Ippoliti, and B.~Vacchini, ``On the energy increase in
  space-collapse models,'' Journal of Physics A: Mathematical and General {\bf
  38} (Aug, 2005) 8017?8038.

\bibitem{Visser:2003vq}
M.~Visser, ``{Jerk and the cosmological equation of state},'' Class. Quant.
  Grav. {\bf 21} (2004) 2603--2616,
\href{http://arXiv.org/abs/gr-qc/0309109}{{\tt arXiv:gr-qc/0309109}}.

\bibitem{Aviles:2012ay}
A.~Aviles, C.~Gruber, O.~Luongo and H.~Quevedo, ``{Cosmography and constraints
  on the equation of state of the Universe in various parametrizations},''
  Phys.\ Rev.\ D {\bf 86} (2012) 123516,
\href{http://arXiv.org/abs/1204.2007}{{\tt arXiv:1204.2007}}.

\bibitem{Bruni:2012sn}
M.~Bruni, R.~Lazkoz and A.~Rozas-Fernandez, ``{Phenomenological models for Unified
  Dark Matter with fast transition},'' Mon.\ Not.\ Roy.\ Astron.\ Soc.\  {\bf 431}
  (2013) 2907,
\href{http://arXiv.org/abs/1210.1880}{{\tt arXiv:1210.1880}}.

\bibitem{Akarsu:2019hmw} 
O.~Akarsu, J.~D.~Barrow, L.~A.~Escamilla and J.~A.~Vazquez, ``{Graduated dark energy:
  observational hints of a spontaneous sign switch in the cosmological constant},''
\href{http://arXiv.org/abs/1912.08751}{{\tt arXiv:1912.08751}}.

\bibitem{Cattoen:2007sk}
C.~Cattoen and M.~Visser, ``{The Hubble series: Convergence properties and
  redshift variables},'' Class. Quant. Grav. {\bf 24} (2007) 5985--5998,
\href{http://arXiv.org/abs/0710.1887}{{\tt arXiv:0710.1887}}.

\bibitem{Dunsby:2015ers}
P.~K.~S.~Dunsby and O.~Luongo, ``{On the theory and applications of modern cosmography},''
  Int.\ J.\ Geom.\ Meth.\ Mod.\ Phys.\  {\bf 13} (2016) 1630002,
\href{http://arXiv.org/abs/1511.06532}{{\tt arXiv:1511.06532}}.

\bibitem{Garcia-Aspeitia:2019yni}
M.~A. Garc\'ia-Aspeitia, C.~Mart\'inez-Robles, A.~Hern\'andez-Almada, J.~Maga{\~n}a, and
  V.~Motta, ``{Cosmic acceleration in unimodular gravity},'' Phys. Rev. {\bf
  D99} (2019) 123525,
\href{http://arXiv.org/abs/1903.06344}{{\tt arXiv:1903.06344}}.

\bibitem{Riess:2017lxs}
A.~G. Riess {\em et al.}, ``{Type Ia Supernova Distances at Redshift $>$ 1.5
  from the Hubble Space Telescope Multi-cycle Treasury Programs: The Early
  Expansion Rate},'' Astrophys. J. {\bf 853} (2018) 126,
\href{http://arXiv.org/abs/1710.00844}{{\tt arXiv:1710.00844}}.

\bibitem{Camarena:2019moy}
D.~Camarena and V.~Marra, ``{Local determination of the Hubble constant and the
  deceleration parameter},''
\href{http://arXiv.org/abs/1906.11814}{{\tt arXiv:1906.11814}}.

\bibitem{Camarena:2019rmj}
D.~Camarena and V.~Marra, ``{A new method to build the (inverse) distance
  ladder},''
\href{http://arXiv.org/abs/1910.14125}{{\tt arXiv:1910.14125}}.

\bibitem{Colgain:2019pck}
E.~\'O.~Colg\'ain, ``{A hint of matter underdensity at low z?},'' JCAP \textbf{09} (2019)
  006,
\href{http://arXiv.org/abs/1903.11743}{{\tt arXiv:1903.11743}}.

\bibitem{Lusso:2019akb}
E.~Lusso, E.~Piedipalumbo, G.~Risaliti, M.~Paolillo, S.~Bisogni, E.~Nardini,
  and L.~Amati, ``{Tension with the flat $\Lambda$CDM model from a
  high-redshift Hubble diagram of supernovae, quasars, and gamma-ray bursts},''
  Astron. Astrophys. {\bf 628} (2019) L4,
\href{http://arXiv.org/abs/1907.07692}{{\tt arXiv:1907.07692}}.

\bibitem{Yang:2019vgk}
T.~Yang, A.~Banerjee, and E.~O. Colg\'ain, ``{On cosmography and flat
  $\Lambda$CDM tensions at high redshift},''
\href{http://arXiv.org/abs/1911.01681}{{\tt arXiv:1911.01681}}.

\bibitem{Velten:2019vwo} 
H.~Velten and S.~Gomes, ``{Is the Hubble diagram of quasars in tension with
  concordance cosmology?},''
\href{http://arXiv.org/abs/1911.11848}{{\tt arXiv:1911.11848}}.

\bibitem{Fukugita:1997bi}
M.~Fukugita, C.~J. Hogan, and P.~J.~E. Peebles, ``{The Cosmic baryon budget},''
  Astrophys. J. {\bf 503} (1998) 518,
\href{http://arXiv.org/abs/astro-ph/9712020}{{\tt arXiv:astro-ph/9712020}}.

\bibitem{Nicastro:2018eam}
F.~Nicastro {\em et al.}, ``{Observations of the MIssing Baryons in the
  warm-hot intergalactic medium},'' \href{http://arXiv.org/abs/1806.08395}{{\tt
  arXiv:1806.08395}}.
[Nature558,406(2018)].

\bibitem{Bregman_2018}
J.~N. Bregman, M.~E. Anderson, M.~J. Miller, E.~Hodges-Kluck, X.~Dai, J.-T. Li,
  Y.~Li, and Z.~Qu, ``The Extended Distribution of Baryons around Galaxies,''
  The Astrophysical Journal {\bf 862} (Jul, 2018) 3.

\bibitem{Garcia-Aspeitia:2019yod}
M.~A. Garc\'ia-Aspeitia, A.~Hern\'andez-Almada, J.~Maga{\~n}a, and V.~Motta, ``{On the
  birth of the cosmological constant and the reionization era},''
\href{http://arXiv.org/abs/1912.07500}{{\tt arXiv:1912.07500}}.

\end{thebibliography}
\end{document}